\documentclass[a4paper,mathptmx,slantedGreek]{aipproc}
\layoutstyle{6x9}
\usepackage{rotating}
\def\SigmaUnit{W m$^{-2}$\,Hz$^{-1}$\,sr$^{-1}$}
\makeatletter
\def\@oddhead{\reset@font{\footnotesize\bf In: \vbox to
0pt{\vss\vtop{\hbox{``High Energy Gamma-Ray Astronomy'', eds.\
F.~Aharonian, W.~Hofmann,}\hbox{F.~Rieger (AIP Conference Proceedings,
Volume 1505, pp5--12), 2012}}}}\hfill\thepage}
\makeatother
\begin{document}

\title{The distribution of SNRs with\\
       Galactocentric radius}

%
%



\author{D.~A.~Green}{address={Cavendish Laboratory, 19 J.~J.~Thomson Ave.,
                              Cambridge, CB3 0HE, U.K.}}

\begin{abstract}
In order to determine the Galactic distribution of supernova remnants (SNRs)
there are two main difficulties: (i) there are selection effects which mean
that catalogues of SNRs are not complete, and (ii) distances are not available
for most SNRs, so distance estimates from the `$\Sigma{-}D$' relation are used.
Here I compare the observed distribution of 69 `bright' SNRs with Galactic
longitude with that expected from the projection of various model
Galactocentric radius distributions. This does not require distances from the
`$\Sigma{-}D$' relation, and selecting only `bright' remnants aims to avoid
major issues with the selection effects. Although this method does not provide
a direct inversion to the 3-D distribution of SNRs in the Galaxy, it does
provide useful constraints on the Galactocentric radius distribution. For a
combined power-law/exponential model for SNR surface density variation with
Galactocentric radius, the best fitted distributions are more concentrated
towards lower radii than the distribution derived by
\citet{1998ApJ...504..761C}.
\end{abstract}

\maketitle
\section{Introduction}

Supernova remnants (SNRs) are important sources of energy and high energy
particles in the Galaxy. Consequently the distribution of SNRs with
Galactocentric radius is of interest for studies of cosmic rays in the Galaxy,
and the high energy $\gamma$-rays they produce from interaction with the
interstellar medium (see, for example, \citet{2012ApJ...750....3A} and
\citet{2012ApJ...752...68V}). Here I discuss some of the problems in
constructing the Galactic distribution of SNRs, particularly due to selection
effects, and the fact that distances are not available for most SNRs. I then
compare the observed distribution of SNRs with Galactic longitude with the
expected distribution from various (simple) models, in order to provide
constraints on the Galactic distribution of SNRs.

\section{Background}

Currently there are 274 catalogued Galactic SNRs
\citet{2009BASI...37...45G}\footnote{See also:
\url{http://www.mrao.cam.ac.uk/surveys/snrs/}}.
In order to obtain the Galactic distribution of SNRs from their observed
Galactic coordinates there are two major hurdles that have to be overcome: (i)
distances are not available for all SNRs, and (ii) there are significant
observational selection effects which means that the catalogue of SNRs in
incomplete.

\subsection{Selection effects}\label{s:selection}

Although some SNRs have first been identified at optical or X-ray wavelengths,
the vast majority have been identified from radio observations, and it is radio
observations -- which are not affected by absorption -- which define the
effective completeness of current SNR catalogues. Basically the selection
effects that apply are (e.g.\ \citet{1991PASP..103..209G}): (i)
surface-brightness ($\Sigma$), and (ii) angular size.

\begin{sidewaysfigure}
\centerline{\includegraphics[angle=270,width=20.0cm]{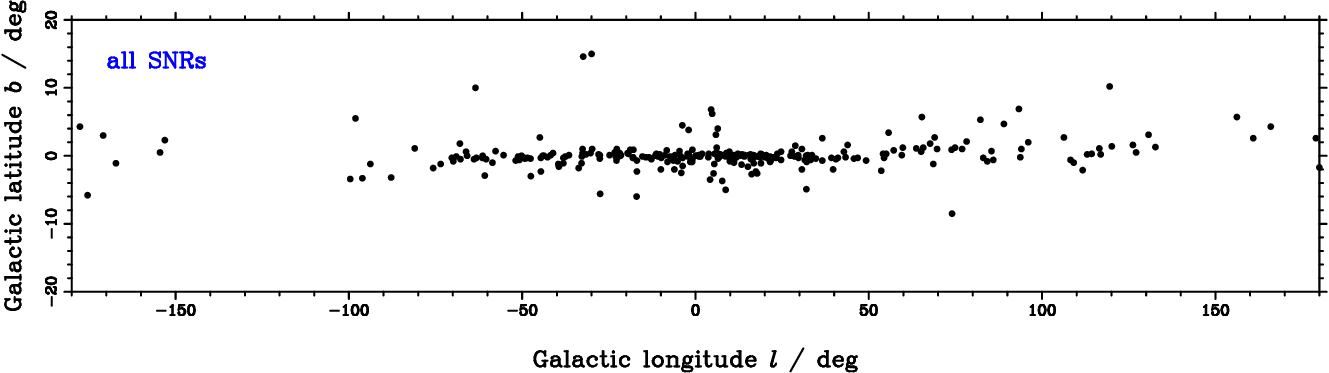}}
\bigskip
\centerline{\includegraphics[angle=270,width=20.0cm]{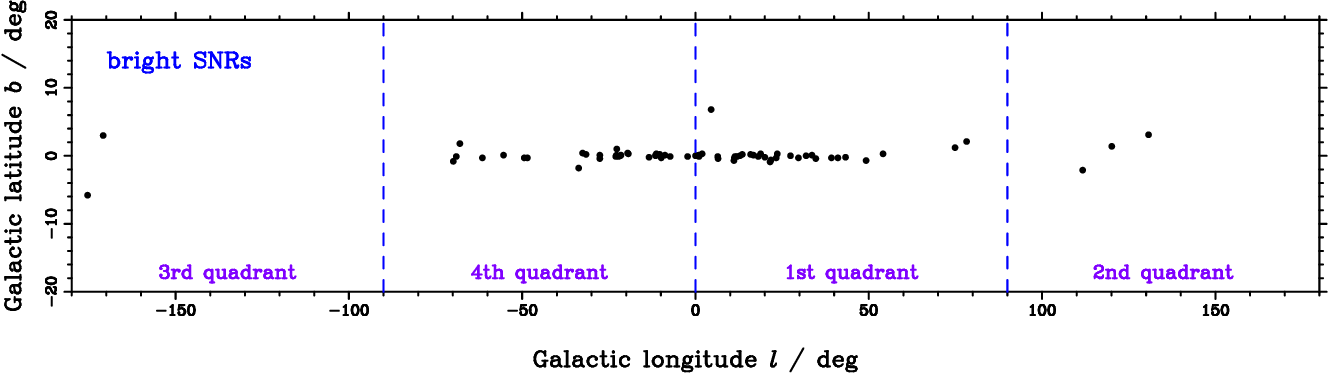}}
\bigskip
\caption{Galactic distribution of: (top) all 274 catalogued SNRs, and (bottom)
the brighter 69 remnants, with surface brightnesses above $10^{-20}$
{\SigmaUnit} at 1~GHz.\label{f:lb}}
\end{sidewaysfigure}

For a SNR to be identified it needs to be bright enough to be distinguished
from the Galactic background. For much of the Galactic plane, the deepest,
large scale survey is that made at 2.7-GHz with the Effelsberg 100-m telescope
(\citet{1990A&AS...85..633R, 1990A&AS...85..691F}), which identified many new
SNRs (\citet{1988srim.conf..293R}). In the region covered by this survey --
i.e.\ $358^\circ < l < 240^\circ$, $|b| < 5^\circ$ -- additional new SNRs have
subsequently been identified, from a variety of other observations. Most of
these newly identified remnants are relatively faint. Of the 60 SNRs that have
been identified from observations made since the Effelsberg 2.7-GHz survey, 54
are below a surface brightness of $10^{-20}$ {\SigmaUnit} at 1~GHz. I take this
surface brightness to be the approximate, single effective $\Sigma$-limit of
the Effelsberg surveys, and hence of the current Galactic SNR catalogue
overall. (Note that the number of catalogued remnants with a surface above this
limit are 35 and 29 in the 1st and 4th Galactic quadrants respectively, which
are consistent within Poisson errors.) In practice, however, it is not easy to
identify SNRs close to this limit in regions of the Galactic plane with high
and complex background radio emission, i.e.\ close to the Galactic Centre. (The
other 6 sources are at most only a factor of about 3 brighter than this nominal
limit, are close to the Galactic Centre, or are $<10$~arcmin in extent, so that
the second selection effect -- discussed below -- also applies.)
Figure~\ref{f:lb} shows the observed distribution in Galactic coordinates of
both (a) all known SNRs, and (b) the 69 SNRs brighter than the nominal surface
brightness completeness limit of $10^{-20}$ {\SigmaUnit} at 1~GHz. Taking this
surface brightness limit shows a distribution more closely correlated towards
both $b=0^\circ$ and the Galactic Centre. This is not surprising, as the lower
radio emission from the Galaxy in the 2nd and 3rd quadrants, and away from
$b=0^\circ$, means it is easier to identify faint SNRs in these regions.

An additional selection effect is that it is generally necessary to resolve a
SNR in order to recognise it structure. Although some limited portions of the
Galactic plane have been observed with high resolutions, available large area
surveys have limited resolution. For example, the Effelsberg surveys noted
above have a resolution of 4.3~arcmin, making it difficult to recognised the
structure of a remnant unless it is $\sim 10$~arcmin or larger in extent. This
means that there is a deficit of small angular size, i.e.\ young but distant
SNRs (see further discussion in \citet{2005MmSAI..76..534G}), but this
selection effect is not easy to quantify. Also, most missing young remnants
will be on the far side of the Galaxy, and appear near $b=0^\circ$ and also
near $l=0^\circ$, which is where the Galactic background is brightest, and
where their is more likely to be confusion with other Galactic source along the
line of sight.

\begin{ltxfigure}
\centerline{\includegraphics[width=10.0cm]{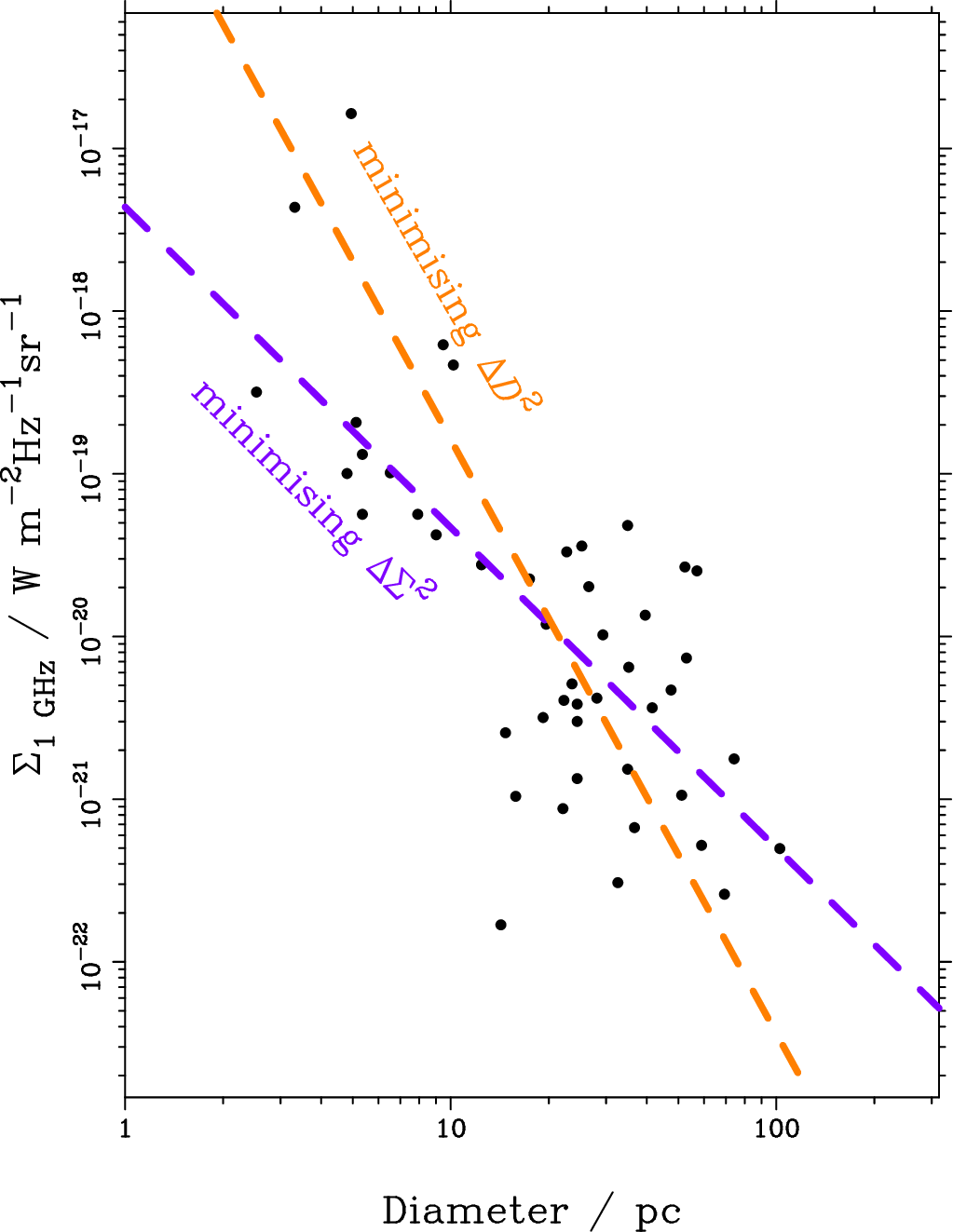}}
\bigskip
\caption{The `$\Sigma{-}D$' -- i.e.\ log(surface brightness, $\Sigma$) versus
log(diameter, $D$) -- for 47 Galactic SNRs for which distance determinations
are available. Also plotted are the least square straight line regressions if
the square deviations in $\Sigma$ or else in $D$ are
minimised.\label{f:sigmad}}
\end{ltxfigure}

\subsection{The `$\Sigma{-}D$' relation}\label{s:sigmad}

In order to directly construct the Galactic distribution of SNRs it is
necessary to know the distance to each catalogued remnant. However, distances
measurements are only available for about 20\% of currently known SNRs. Often
the statistical correlation between the \emph{observed} surface brightness
($\Sigma$) and diameter, $D$, has been used to derive diameters -- and hence
distances, $d=D/\theta$, using the \emph{observed} angular size -- for SNRs for
which distance determinations are not available. This is using the
`$\Sigma{-}D$' relation, usually parameterised as
$$
  \Sigma = C D^{-n}
$$
since, for remnants with known distances, physically small ones tend to have
larger surface brightnesses than larger remnants. (Note that, as discussed in
\citet{2005MmSAI..76..534G}, much of this correlation is arguably due to a
$D^{-2}$ bias.) In practice, however, SNRs show a wide range of physical
diameters for a given surface brightness, and so the `$\Sigma{-}D$' relation is
of limited use for determining distances to individual remnants (e.g.\
\citet{1991PASP..103..209G, 2005MmSAI..76..534G}). And because of the selection
effects noted above, the full range of properties of SNRs may be even larger
than is evident currently, since faint, small angular size (i.e.\ likely
physically small also) remnants are difficult to identify.

Moreover, as noted by \citet{2005MmSAI..76..534G}, some use of the
`$\Sigma{-}D$' relation for statistical studies has been affected by using
inappropriate straight line regressions. If the $\Sigma{-}D$ relation is to be
used to predict $D$-values from observed $\Sigma$-values, then a least square
regression that minimises the square deviations in $D$ should be used, not one
that minimises the square deviations in $\Sigma$. Given there is a quite a
large scatter in the $\Sigma{-}D$ plane for SNRs with known distances, the
differences between these regressions are significant (e.g.\ see
\citet{1990ApJ...364..104I}), particularly for the fainter or brighter
remnants. For example, \citet{1998ApJ...504..761C} minimised the square
deviations on $\Sigma$, and obtained a $\Sigma{-}D$ relation with $n = 2.64 \pm
0.26$ (for 37 `shell' remnants, including Cas~A), which they commented was a
considerably flatter $\Sigma{-}D$ slope than had been obtained previously
(e.g.\ \citet{1979AuJPh..32...83M} derived $n=3.8$, \citet{1981SvAL....7...17L}
derived $n=3.45$). However, by minimising square deviations in $D$, for the
calibrators used by \cite{1998ApJ...504..761C}, $n = 3.53 \pm 0.33$ is
obtained, which is a much steeper $\Sigma{-}D$ slope, and is consistent with
previously derived slopes. This is illustrated in Figure~\ref{f:sigmad}, for a
sample of SNRs with known distances (these are not the same sample as used by
\citet{1998ApJ...504..761C}, but show the main feature, the difference in
fitted slopes for different regressions). \citet{1998ApJ...504..761C} went on
to derive the Galactic distribution of SNRs, using their $\Sigma{-}D$ slope,
which is not steep enough, and hence \emph{overestimates} the diameters -- and
distances -- for the majority of SNRs, which are relatively faint.

\begin{ltxfigure}
\centerline{\raisebox{-\height}{(a)}
            \includegraphics[angle=270,width=14cm,clip=]{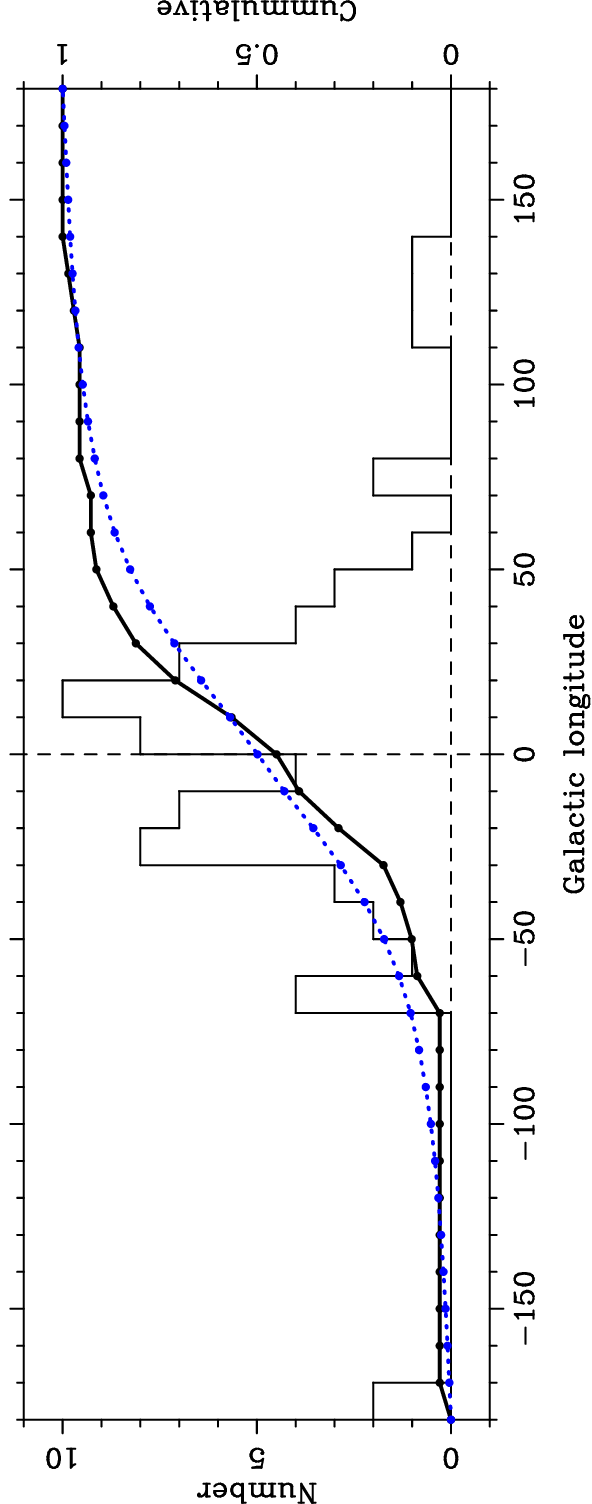}}
\bigskip
\centerline{\raisebox{-\height}{(b)}
            \includegraphics[angle=270,width=14cm,clip=]{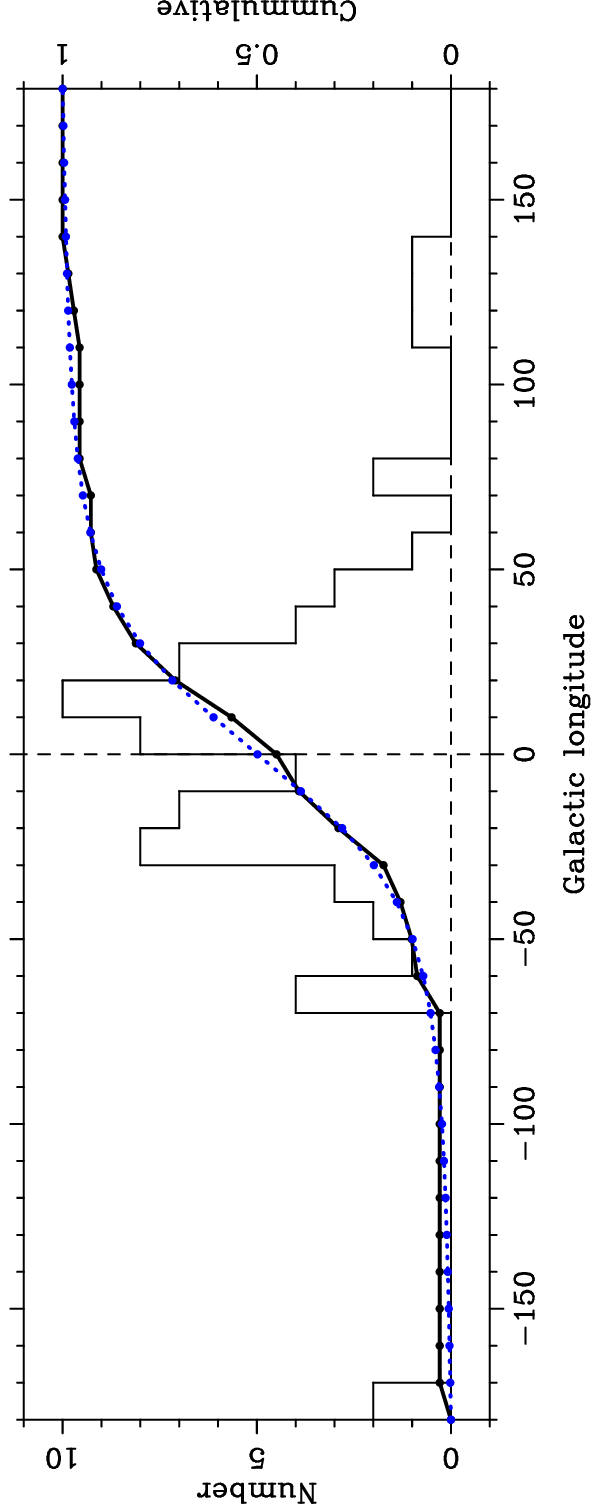}}
\bigskip
\centerline{\raisebox{-\height}{(c)}
            \includegraphics[angle=270,width=14cm,clip=]{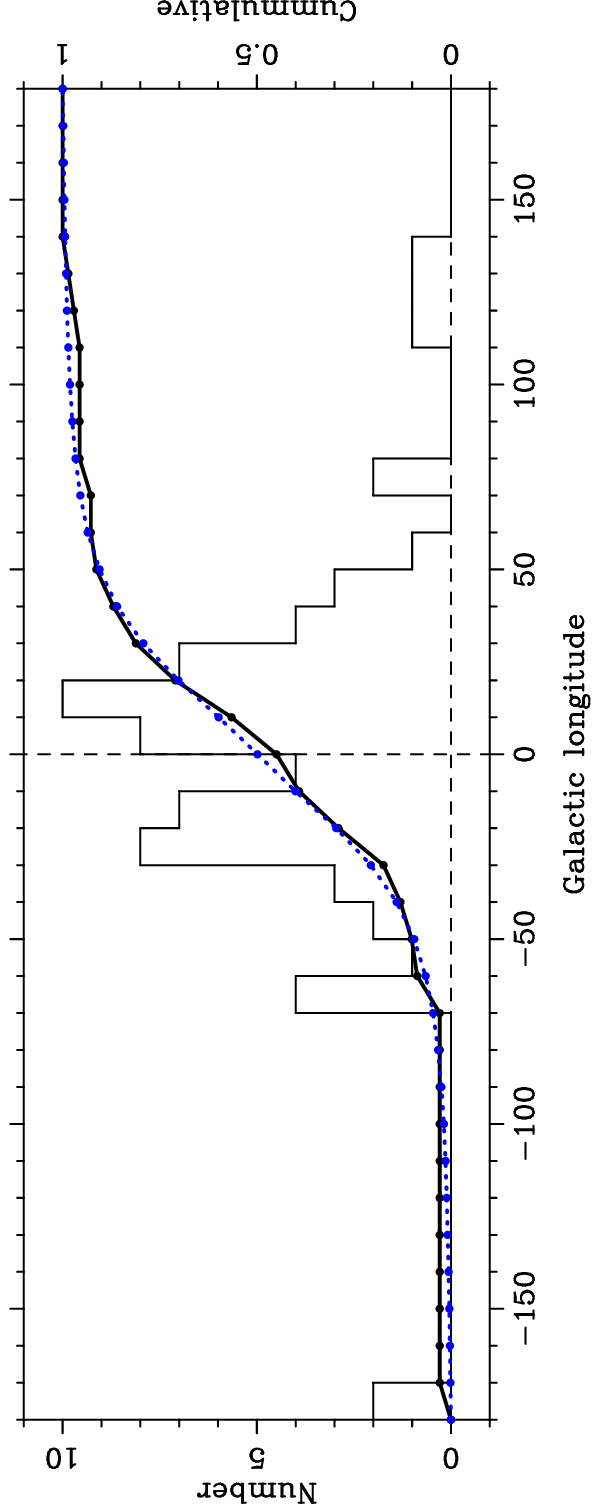}}
\bigskip
\caption{Observed $l$-distribution of bright Galactic SNRs, plotted as
histogram (left scale), and cumulative fraction (solid line, right scale), plus
cumulative fraction for a model distribution (dotted line, right scale). The
three models are for the surface density of SNRs varying with Galactocentric
radius, $R$, as (a) $\propto ({R}/{R_\odot})^{2.0} \exp \left[
-3.5(R-R_\odot)/R_\odot \right]$ (as derived by \citet{1998ApJ...504..761C}),
(b) $\propto ({R}/{R_\odot})^{0.8} \exp \left[ -3.5(R-R_\odot)/R_\odot
\right]$, and (c) $\propto ({R}/{R_\odot})^{2.0} \exp \left[
-5.1(R-R_\odot)/R_\odot \right]$.}\label{f:sd}
\end{ltxfigure}

\section{The Galactic distribution of SNRs}\label{s:gd}

Rather than attempting to derive the 3-D distribution of Galactic SNRs from
their observed properties, which would require reliable distances -- and
corrections for selection effects if all known SNRs were to be used -- I follow
the method used in \citet{1996ssr..conf..341G, 2004BASI...32..335G}. This
simply compares the observed distribution of SNRs in Galactic longitude, $l$,
using only brighter remnants above the nominal surface-brightness limit of
current catalogues, with the expected distribution projected from various
models.

\begin{ltxfigure}
\centerline{\includegraphics[angle=270,width=13.5cm]{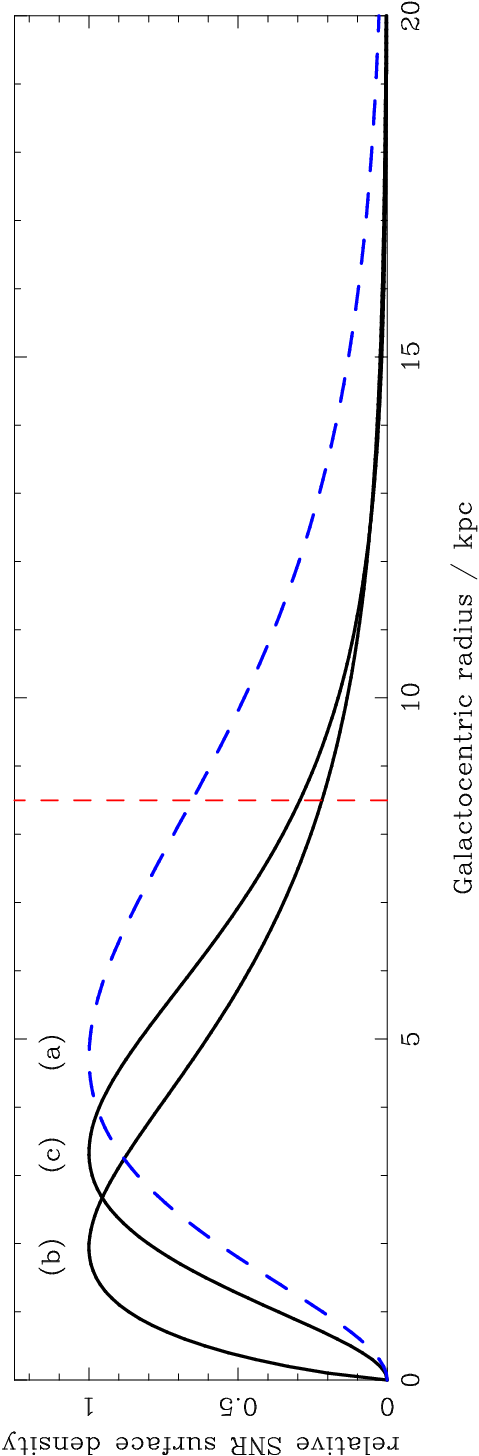}}
\bigskip
\caption{Variation of the surface density of SNRs with Galactocentric radius,
$R$, for the power-law/exponential models shown in Figure~\ref{f:sd} and
discussed in Section~\ref{s:gd}: dashed line for \citet{1998ApJ...504..761C}'s
distribution (a), and solid lines for models (b) and (c).}\label{f:radial}
\end{ltxfigure}

In \citet{1996ssr..conf..341G, 2004BASI...32..335G} a simple Gaussian model for
the surface density of SNRs in the Galactic disk was used, which peaks at the
Galactic Centre. However, the observed radial distributions of other star
formation tracers (e.g.\ pulsars and star formation regions, see
\citet{2000A&A...358..521B, 2006MNRAS.372..777L}) indicate a minimum at the
Galactic Centre, and so here I use a two parameter mixed power-law/exponential
radial distribution for the density of SNRs with Galactocentric radius, $R$,
i.e.
$$
  \propto \left( \frac{R}{R_\odot} \right)^A
  \exp \left[ -B \frac{(R-R_\odot)}{R_\odot} \right]
$$
with $R_\odot = 8.5$~kpc, the distance to the Galactic Centre.
Figure~\ref{f:sd} shows the observed distribution of SNRs with $\Sigma >
10^{-20}$ {\SigmaUnit} from \citet{2009BASI...37...45G}, together with
projected model distributions for: (a) $A=2.0$, $B=3.6$, (b) $=0.8$, $B=3.5$,
and (c) $=2.0$, $B=5.1$. The model distribution shown in Figure~\ref{f:sd}(a),
which corresponds to the power-law/exponential distribution obtained by
\citet{1998ApJ...504..761C}, is clearly broader than the observed distribution
of `bright' SNRs above the nominal surface brightness limit of current SNR
catalogues (which is to be expected, given the systematic difference due to the
regression used by \citet{1998ApJ...504..761C} noted in
Section~\ref{s:sigmad}). The model distributions shown in Figure~\ref{f:sd}(b)
and (c) are the best fit -- in the sense of minimum sum of least squares of the
difference between the observed and model cumulative distributions --
power-law/exponential radial distribution varying parameters $A$ and $B$
respectively, but keeping the other parameter as derived by
\citet{1998ApJ...504..761C}. Models (b) and (c) have similar least squares
differences from the observed cumulative distribution, but correspond to
somewhat different distributions in Galactocentric radius, as shown in
Figure~\ref{f:radial}. This shows that there is degeneracy between the
parameters $A$ and $B$ in the power-law/exponential distribution model. As
discussed above, any residual selection effects that apply to the sample of
`bright' SNRs will mean that it is more likely that remnants closer to
$l=0^\circ$ may have been missed. This means that the true distribution may be
more concentrated towards $l=0^\circ$, and hence the distribution will be more
concentrated towards smaller Galactocentric radii.

\section{Conclusions}

Observational selection effects, and the lack of distances for many Galactic
SNRs means that deriving the Galactic SNR distribution is not straightforward.
For a sample of 69 `bright' SNRs -- i.e.\ those not strongly affected by
selection effects -- the observed $l$-distribution is shown not to be
consistent with the Galactic distribution with Galactocentric radius derived by
\citet{1998ApJ...504..761C}.

\begin{theacknowledgments}

I thank Irina Stefan for useful comments on early drafts of this paper.

\end{theacknowledgments}


\end{document}